\documentclass{emulateapj}
\def\pr{^{\prime}}
\def\2pr{^{\prime \prime}}

\def\greatsim{\mathrel{\raise.3ex\hbox{$>$\kern-.75em\lower1ex\hbox{$\sim$}}}}
\def\lesssim{\mathrel{\raise.3ex\hbox{$<$\kern-.75em\lower1ex\hbox{$\sim$}}}}

\slugcomment{Submitted to {\em Astrophysical Journal Letters}}

\shorttitle{Strong Lensing Arcs in Cluster at z=1.026}
\shortauthors{Huang et al.}

\begin{document}


\title{HST\footnotemark[1] Discovery of a \textit{z}=3.9 Multiply Imaged Galaxy Behind the Complex
Cluster Lens WARPS J1415.1+36 at \textit{z}=1.026}

\author{
X.~Huang\altaffilmark{2},
T.~Morokuma\altaffilmark{3,4},
H.~K.~Fakhouri\altaffilmark{2,5},
G.~Aldering\altaffilmark{5},
R.~Amanullah\altaffilmark{6},
K.~Barbary\altaffilmark{2,5},
M.~Brodwin\altaffilmark{7, 8},
N.~V. Connolly\altaffilmark{9}, 
K.~S.~Dawson\altaffilmark{10},
M.~Doi\altaffilmark{11},
L.~Faccioli\altaffilmark{5},
V.~Fadeyev\altaffilmark{12},
A.~S.~Fruchter\altaffilmark{13}, 
G.~Goldhaber\altaffilmark{2,5},
M.~D.~Gladders\altaffilmark{14},
J.~F.~Hennawi\altaffilmark{2},
Y.~Ihara\altaffilmark{3,11},
M.~J.~Jee\altaffilmark{15},
M.~Kowalski\altaffilmark{16},
K.~Konishi\altaffilmark{17},
C.~Lidman\altaffilmark{18},
J.~Meyers\altaffilmark{2,5},
L.~A.~Moustakas\altaffilmark{19},
S.~Perlmutter\altaffilmark{2,5},
D.~Rubin\altaffilmark{2,5},
D.~J.~Schlegel\altaffilmark{5},
A.~L.~Spadafora\altaffilmark{5},
N.~Suzuki\altaffilmark{5},
N.~Takanashi\altaffilmark{4},
N.~Yasuda\altaffilmark{17}
}

\footnotetext[1]{Based on observations made with
the NASA/ESA Hubble Space Telescope, obtained from the data
archive at the Space Telescope Institute. STScI is operated by the association of Universities for
Research in Astronomy, Inc. under the NASA contract NAS 5-26555.
The observations are associated with program 10496.}

\altaffiltext{2}{Department of Physics, University of California Berkeley, Berkeley, CA 94720}
\altaffiltext{3}{Research Fellow of the Japan Society for the Promotion of Science}
\altaffiltext{4}{National Astronomical Observatory of Japan, 2-21-1 Osawa, Mitaka, Tokyo,181-8588, Japan}
\altaffiltext{5}{E.O. Lawrence Berkeley National Laboratory, 1 Cyclotron Rd., Berkeley CA, 94720}
\altaffiltext{6}{ Department of Physics, Stockholm University, Albanova University Center, S-106 91 Stockholm, Sweden}
\altaffiltext{7} {Harvard-Smithsonian Center for Astrophysics, 60 Garden Street, Cambridge, MA 02138, USA}
\altaffiltext{8} {W. M. Keck Postdoctoral Fellow at the Harvard-Smithsonian Center for Astrophysics}
\altaffiltext{9} {Department of Physics, Hamilton College, Clinton, NY 13323, USA}
\altaffiltext{10}{Department of Physics and Astronomy, University of Utah, Salt Lake City, UT 84112}
\altaffiltext{11}{Institute of Astronomy, Graduate School of Science, University of Tokyo 2-21-1 Osawa, 
Mitaka, Tokyo 181-0015, Japan}
\altaffiltext{12}{Santa Cruz Institute for Particle Physics, University of California, Santa Cruz, CA 95064, USA}
\altaffiltext{13}{Space Telescope Science Institute, 3700 San Martin Drive, Baltimore, MD 21218, USA}
\altaffiltext{14}{Department of Astronomy and Astrophyics, The University of Chicago, 5640 S. Ellis Ave, 
Chicago, IL 60637}
\altaffiltext{15}{Department of Physics, University of California, Davis, One Shields Avenue, 
Davis, CA 95616, USA}
\altaffiltext{16}{Humboldt Universit¨at Institut f¨ur Physik, Newtonstrasse 15, 12489 Berlin, Germany}
\altaffiltext{17}{Institute for Cosmic Ray Research, University of Tokyo, 5-1-5, Kashiwanoha, Kashiwa, Chiba,
277-8582, Japan}
\altaffiltext{18}{Oskar Klein Centre, Department of Physics, Stockholm University, Roslagstullsbacken 21, 106 91 Stockholm, Sweden}
\altaffiltext{19}{Jet Propulsion Laboratory, California Institute of Technology, 4800 Oak Grove Dr, MS 169-327, Pasadena, CA 91109}

\email{xhuang@lbl.gov}

\begin{abstract}
We report the discovery of a multiply lensed Ly$\alpha$ emitter at \textit{z}=3.90
behind the massive galaxy cluster WARPS J1415.1+3612 at \textit{z}=1.026.  
Images taken by the \textit{Hubble Space Telescope} (\textit{HST}) using
ACS reveal a complex lensing system that produces
a prominent, highly magnified arc and a triplet of smaller arcs grouped tightly around 
a spectroscopically confirmed cluster member.
Spectroscopic observations using FOCAS on Subaru confirm strong Ly$\alpha$ emission in the 
source galaxy and provide redshifts for more than 21 cluster members, from which we obtain
a velocity dispersion of $807 \pm 185\ \rm km\ s^{-1}$.
Assuming a singular isothermal sphere profile, the mass within the Einstein ring 
($7.13 \pm 0.38\2pr$) corresponds to a central velocity dispersion of 
$686^{+15}_{-19}\ \rm km\ s^{-1}$ for the cluster, consistent with the value 
estimated from cluster member redshifts.  Our mass profile estimate from 
combining strong lensing and dynamical analyses is in 
good agreement with both X-ray and weak lensing results.
\end{abstract}

\section{Introduction}\label{sec:intro}


Surveys like SpARCS \citep{wilson08a}, SPT \citep{ruhl04a},
the Red-Sequence Cluster Surveys 
(RCS, \citealt{gladders04a} and RCS-2, \citealt{gladders05a, yee07a})
and the IRAC Shallow Survey \citep{eisenhardt08a}
will discover thousands of galaxy clusters out to redshifts beyond $z=1$.
These large surveys will be used to probe dark energy and the underlying
cosmology through direct measurements of the evolution of the cluster mass function.
To fully utilize these data it will be essential to derive reliable
mass estimates from the cluster properties measured in these surveys.

Great effort has been dedicated to the development of cluster mass-observable
relations in targeted cluster observations.
Sunyaev-Zel'dovich (SZ) and X-ray observations probe the hot ionized intracluster gas
and have been used to derive masses
\citep{laroque06a, allen08a} and scaling relations that tie mass
to the observed SZ quantities \citep{bonamente08a}.
\citet{hicks06a} derive scaling relations between observed X-ray
properties and the optical richness of clusters while \citet{evrard08a} model
the relationship between velocity dispersion of cluster galaxies and cluster mass.
Weak lensing measurements have been used to calibrate optical richness measurements from a large sample
of clusters from the Sloan Digital Sky Survey \citep{koester07a,johnston07a}
and to calibrate X-ray measurements through the 
mass-temperature relations \citep{hoekstra07a, bardeau07a, berge08a}.
Strong lensing measurements typically have much smaller errors compared with the 
other methods but are restricted to the central region of clusters and are rare.
Nevertheless, strong lensing systems have been studied in small samples
of clusters \citep{comerford06a, hennawi08a}.

Large cluster surveys out to $z$ $\sim$ 1 have the potential to constrain the dark energy
equation of state \citep[e.g.][]{voit05a}, provided that the mass-observable 
relations are well calibrated.
Full exploitation of the more distant clusters requires deep space-based observations.
As a result, mass-observable relations have been studied in
far fewer clusters at $z>1$ relative to low redshift clusters.
In a 219 orbit program (Program number 10496, PI: Perlmutter) to search for
supernovae (SNe) with the Hubble Space Telescope (HST),
we observed 25 of the highest redshift galaxy clusters known at the time 
\citep{dawson09a}.
These images support a rich program of cluster studies including the calibration
of mass proxies at $z>1$.
One particular cluster from this program, selected from the Wide Angle \textit{ROSAT}
Pointed Survey, WARPS J1415.1+3612 at $z=1.026$ \citep{perlman02a},
has already been studied in X-ray \citep{maughan06a, allen08a} and SZ \citep{muchovej07a} observations.
We now have deep images from the Advance Camera for Surveys (ACS)
and spectroscopy from the Faint Object Camera and Spectrograph \citep[FOCAS:][]{kashikawa02a} on Subaru
that eveal a pronounced strong lensing arc of a
source Ly$\alpha$ emitting galaxy at $z=3.90$.  
The new data from this program combined with previous and ongoing measurements enable 
a multi-probe analysis of the cluster mass-observable relation for this high redshift cluster.

Here we present an analysis of the strong lensing and dynamical mass 
of WARPS J1415.1+3612.
The letter is organized as follows:  we describe the observations and data in
\S\ref{sec:observations},
mass estimates are derived and compared in \S\ref{sec:discussion},
and the summary is found in \S\ref{sec:summary}. 
Throughout this letter we use AB magnitudes and assume a cosmology with
$\Omega_M = 0.3$, $\Omega_\Lambda = 0.7$, and $h = 0.7$.


\section{Observations}\label{sec:observations}


\subsection{ACS Images}\label{subsec:acs}

WARPS J1415.1+3612 was observed 7 times with ACS from November 2005 through April 2006
for a total integration of 2425 seconds in the F775W filter and
9920 seconds in the F850LP filter, hereafter $i_{775}$ and $z_{850}$.
Individual exposures were coadded using MultiDrizzle \citep{fruchter02a} at a
resolution of $0.05\2pr /$pixel.  We use 
25.678 and 24.867 as zeropoints for $i_{775}$ and $z_{850}$ respectively \citep{sirianni05a}.

\begin{figure}[!h]
\begin{center}
\includegraphics[scale=0.45]{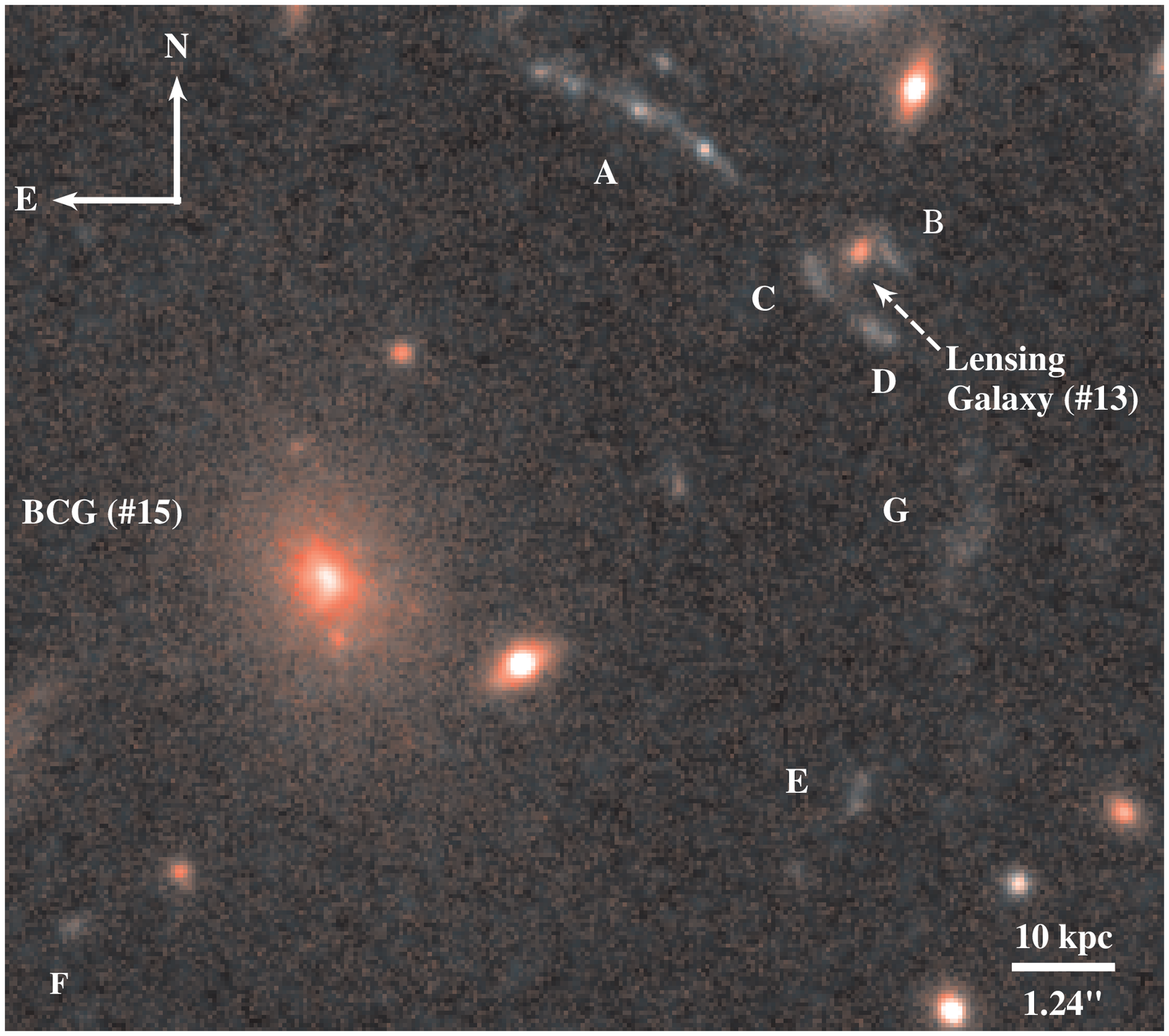}
\includegraphics[scale=0.45]{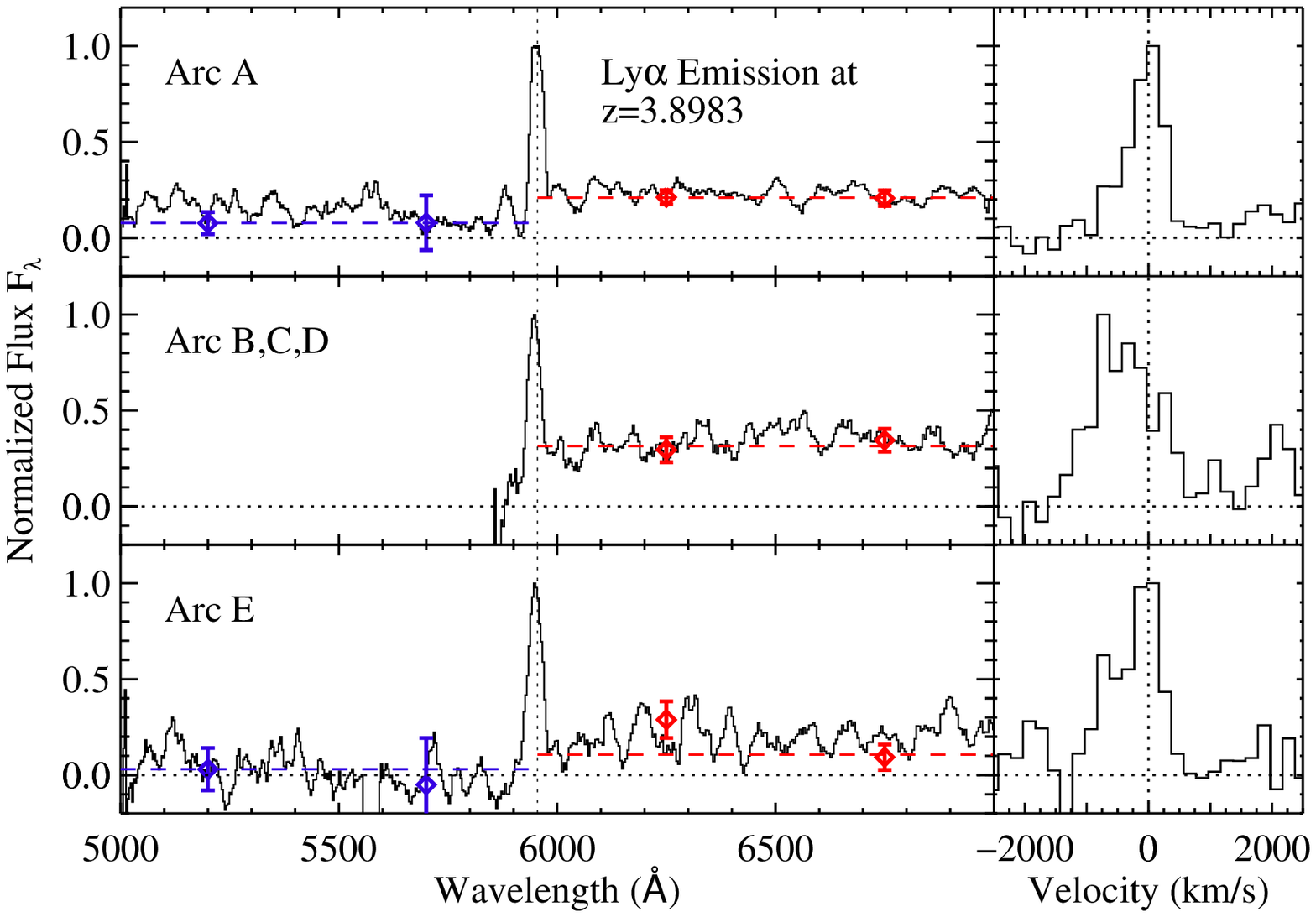}
\end{center}
\caption[1-D Spectrum of Arc System]{\label{fig:allspect}
{\bf Top:}
Central Region of WARPS J1415.1+3612.  Strong lensing arcs are labeled ``A"-``E"
and cluster galaxies of interest are labeled ``\#13" and ``\#15".
The radius of the Einstein ring is taken to be the average of the radii of 
two imaginary circles (not shown), both centered on the BCG.  The first circle with radius 
$6.75\2pr$ passes through the center of arc A and the second, with radius $7.51\2pr$, 
passes through cluster member \#13 (the lensing galaxy).  
F and G are discussed in the text.
{\bf Bottom:}
The panel on the left 
shows smoothed spectra of arc A; the triplet B, C and D; and arc E. 
Zero flux is shown as the horizontal dotted lines.  The diamonds with error bars are the 
average values of flux in 500 \AA\ bins.
The differences between the averages of the flux for wavelengths longer and shorter than the peak for
arcs A and E (dashed lines) indicates the detection of 
discontinuity around the emission feature.  
The panel on the right shows the same spectra smoothed to the spectrograph resolution 
(FWHM=600 $\rm km\ s^{-1}$) and resampled by 200 $\rm km\ s^{-1}$ pixel in the log linear scale.
}
\end{figure}

The deep ACS images revealed a complex strong lensing system near the cluster core.
The composite color image from the $i_{775}$ and $z_{850}$ data is shown in Figure~\ref{fig:allspect}.
The arc system consists of at least five images.
The main arc (A) is $6.75\2pr$ (54.4 kpc at the cluster redshift) from the
center of the cluster, which we take to be the position of the brightest cluster 
galaxy (BCG).
To the SW of the main arc,
there is a triplet of arcs (B, C and D) around a spectroscopically confirmed cluster member (\#13).
About $5\2pr$ south of the triplet, there is a fifth arc (E).  The total $i_{775}$ isophotal 
magnitudes for  the arcs 
A - E are 23.44,  25.75, 25.46, 25.13 and 25.62, respectively.  We use the $i_{775}$ magnitudes
because they suffer less contamination than the $z_{850}$ magnitudes from the much redder 
elliptical cluster members.  The magnitudes for the arcs B, C and D are 
obtained after the lensing galaxy has been subtracted using the b-spline model
of \citet{bolton06a}. 
These five images lie very close to an imaginary arc centered on the BCG that subtends 
slightly more than \(90^\circ\).  The radius of the Einstein ring for the 
cluster potential is taken to be the average of the two circles mentioned in 
Figure~\ref{fig:allspect}, $7.13\pm0.38\2pr$.  The uncertainty in the radius is 
determined from the difference 
between the two circles and the average. 

The $B$ band luminosity for this cluster is estimated by summing
up light from all the galaxies within approximately 1 Mpc of the
BCG and then subtracting a background we estimate from the GOODS 
\citep{giavalisco04a} images.  We apply a $K$-correction to transform 
$z_{850}$ to restframe $B$ magnitude.  The total luminosity is determined 
to be $L_{B} = 2.92 \pm 0.88 \times 10^{12} L_{B\odot}$.


\subsection{Subaru Spectroscopy}\label{subsec:spec}

Spectroscopic observations were made with the FOCAS spectrograph
on the Subaru telescope.
SN candidates provided the primary targets for spectroscopic observations.
The strong lensing arcs described in \S\ref{subsec:acs} and galaxies with $i_{775}$-$z_{850}$
color consistent with early type galaxies at the cluster redshift
provided secondary targets for multi-slit spectroscopy.  We obtained spectra for a total of 
138 objects and measured redshifts for 95 of them.  Six masks were used, four with the 
300R/SO58 grism and filter and two with the 300B/SY47 grism and filter. The total integration 
time is approximately two hours for each mask.  Arcs B, C and D were observed
with the former setup, resulting in spectra that start at 5800\AA, while
arc E was observed with the latter, resulting in spectra that
cover a bluer wavelength range.
Arc A was observed with both setups.
The slit width is $0.8\2pr$, corresponding to a spectral resolution of $R \sim 500$.
Further details can be found in Morokuma et al.\ 2009 (in preparation).

Redshifts for cluster members are obtained by fitting a series of galaxy 
spectral eigenfunctions to the spectra weighted by the error spectra.
For the majority of the spectra, the 3933 \AA\ and 3968 \AA\ Ca II H and K absorption lines
are the dominant feature detected, along
with the 4000 \AA\ break and the [OII] emission line.  Each redshift is assigned a
quality flag $Q$, with $Q=3$ for those with multiple strong features,
$Q=2$ for those with a single strong feature and $Q=1$ for those with only weakly detected features.  
Cluster members selected from the spectroscopically confirmed galaxies inside a radius of $\sim$ $2.1\pr$ 
are listed in Table 1.  The redshift of the cluster is determined to be 1.026
from the average of the cluster member redshifts listed in the table.
At this redshift, a radius of $2.1\pr$ corresponds to approximatly 1 Mpc.
We calculate the cluster velocity dispersion from the 21 galaxies with $Q>1$.
Using the biweight scale estimator of \citet{beers90a},
we estimate $\sigma_{\rm v} = 807 \pm 185\ \rm km\ s^{-1}$  in the cluster rest frame.
The measurement uncertainty is
determined from bootstrap resampling of the data.  

\begin{deluxetable}{lcccl}
\tablewidth{0pt}
\tabletypesize{\scriptsize}
\centering
\tablecaption{\label{tab:members} Spectroscopic redshifts}
\tablehead{\colhead{Object ID} & \colhead{R.A. (J2000)} & \colhead{Dec. (J2000)} & \colhead{\textit{z}} & \colhead{\textit{Q}} }
\startdata
1 &  14:15:13.08 & +36:12:25.4 &  1.0262 $\pm$ 0.0003  &   3   \\ 
2 &  14:15:12.68 & +36:12:59.6 &  1.0331 $\pm$ 0.0003  &   3   \\ 
3 &  14:15:11.27 & +36:14:35.7 &  1.033  $\pm$ 0.001   &   2   \\ 
4 &  14:15:11.64 & +36:11:37.8 &  1.028  $\pm$ 0.002   &   1   \\ 
5 &  14:15:07.08 & +36:14:15.6 &  1.0255 $\pm$ 0.0003  &   3   \\ 
6 &  14:15:07.68 & +36:13:06.3 &  1.025  $\pm$ 0.001   &   2   \\ 
7 &  14:15:13.95 & +36:12:37.8 &  1.017  $\pm$ 0.001   &   2   \\ 
8 &  14:15:10.36 & +36:11:31.2 &  1.0256 $\pm$ 0.0003  &   3   \\ 
9 &  14:15:15.91 & +36:12:58.9 &  1.0187 $\pm$ 0.0003  &   3   \\ 
10 & 14:15:10.66 & +36:11:39.8 &  1.028  $\pm$ 0.001   &   2   \\ 
11 & 14:15:08.08 & +36:12:21.0 &  1.0234 $\pm$ 0.0003  &   3   \\ 
12 & 14:15:10.54 & +36:12:09.9 &  1.0264 $\pm$ 0.0003  &   3   \\ 
13$^a$ & 14:15:10.59 & +36:12:07.9 &  1.023  $\pm$ 0.002   &   1   \\ 
14 & 14:15:10.19 & +36:11:49.6 &  1.0313 $\pm$ 0.0003  &   3   \\ 
15$^a$ & 14:15:11.12 & +36:12:04.0 &  1.0252 $\pm$ 0.0003  &   3   \\
16 & 14:15:11.47 & +36:12:22.8 &  1.018  $\pm$ 0.001   &   2   \\
17 & 14:15:06.61 & +36:12:18.0 &  1.033  $\pm$ 0.001   &   2   \\
18 & 14:15:09.12 & +36:13:09.1 &  1.0260 $\pm$ 0.0003  &   3   \\
19 & 14:15:13.45 & +36:12:56.9 &  1.035  $\pm$ 0.002   &   1   \\
20 & 14:15:12.60 & +36:11:09.7 &  1.023  $\pm$ 0.001   &   2   \\
21 & 14:15:00.02 & +36:13:02.8 &  1.011  $\pm$ 0.001   &   2   \\
22 & 14:15:05.89 & +36:13:03.1 &  1.025  $\pm$ 0.001   &   2   \\
23 & 14:15:13.00 & +36:13:07.0 &  1.0301 $\pm$ 0.0003  &   3   \\
24 & 14:15:08.41 & +36:13:04.6 &  1.035  $\pm$ 0.002   &   1   \\
25 & 14:15:10.82 & +36:10:33.0 &  1.023  $\pm$ 0.001   &   2   \\
\\
A & 14:15:10.80 & +36:12:09.1  &   3.8983 $\pm$ 0.0005   &   2     \\
B-D$^b$ & 14:15:10.57 & +36:12:06.9 &  3.891 $\pm$ 0.002   &   2   \\
E & 14:15:10.58 & +36:12:00.8 &  3.897 $\pm$ 0.001   &   2         
\enddata
\tablenotetext{a}{Highlighted in Figure~1}
\tablenotetext{b}{Triplet unresolved in 2-d spectra.}  
\end{deluxetable}

Spectra from each of the arc candidates (A - E) show significant emission around 5950~\AA\
(Figure~\ref{fig:allspect}).  The consistency in the observed emission wavelengths, 
surface brightness, color, and morphology of these five objects, and their arrangement 
around a circle centered on the BCG all strongly suggest that this is a multiply imaged 
system at $z>1.026$.  In this case, the line is almost certainly
Ly$\alpha$ at restframe 1215.67 \AA.  The 
detection of discontinuity in flux for wavelengths 
bluer and redder than the emission feature in the spectra for arcs A and E  
lends support to the identification of this line with Ly$\alpha$.
Assuming this emission is due to Ly$\alpha$,
the redshifts of each arc are reported in Table~\ref{tab:members}.
As indicated by the apparent discrepancy in redshifts, there is
$\sim9$ \AA\ (or $453\ \rm km\ s^{-1}$ in the rest frame of the cluster, 
Figure~\ref{fig:allspect}) difference between the observed wavelengths of Ly$\alpha$
in arc A and in the triplet (arcs B, C, and D).  This is 
3 times the statistical error, which takes into account the spectral resolution. 
This difference is likely due to the placement of the arcs B, C and D
near the edges of the slit and contamination from the lensing galaxy
(cluster member \#13).

Preliminary spectroscopic results show that object F in Figure~\ref{fig:allspect} is a possible 
counter arc of the same source galaxy.  Object G in Figure~\ref{fig:allspect} has the 
same color as arcs A - E and lies near the Einstein ring but spectrocopic data 
are inconclusive as to whether it has the same redshift as the confirmed arcs.


\section{Discussion}\label{sec:discussion}

The total mass inside the Einstein ring of $r_{E}=57.5 \pm 3.1$ kpc is 
\(1.96^{+0.22}_{-0.20}\times10^{13} M_{\odot}\)
 \citep[e.g.][]{narayan96a}.  
If we assume a singular isothermal sphere (SIS) profile for the central region of the cluster, 
this mass corresponds to a rest frame velocity dispersion of 
$686^{+15}_{-19}\ \rm km\ s^{-1}$, consistent with the value estimated from cluster member redshifts.  
We can estimate the magnification factor for arc A by using $\mu = |1/(1-r_E/r)|$ 
\citep[e.g.][]{narayan96a} for the SIS profile, where $r$ is the distance of the image from the lens center.  
If we take $r_E = 7.13 \2pr$ (or 57.5 kpc), $\mu \sim 18$ for arc A.  
This implies that the unmagnified $i_{775}$ magnitude for arc A is $\sim$ 26.6.

To get a qualitative estimate of the change in the derived strong lensing
mass caused by relaxing our assumption of a spherically symmetric mass
distribution, we construct elliptical NFW models \citep{glose02a} with a
critical curve that passes through at least part of arcs A and E and passes
between arc C and the lensing galaxy (cluster member \#13).  The model with the
largest ellipticity that still meets these requirements has an ellipticity of 0.1.  
The mass enclosed within its critical curve is $1.79\times 10^{13} M_{\odot}$, 
within the errors of the spherical models.

\begin{figure}[h!]  
\begin{center}
\includegraphics[width=0.5\textwidth]{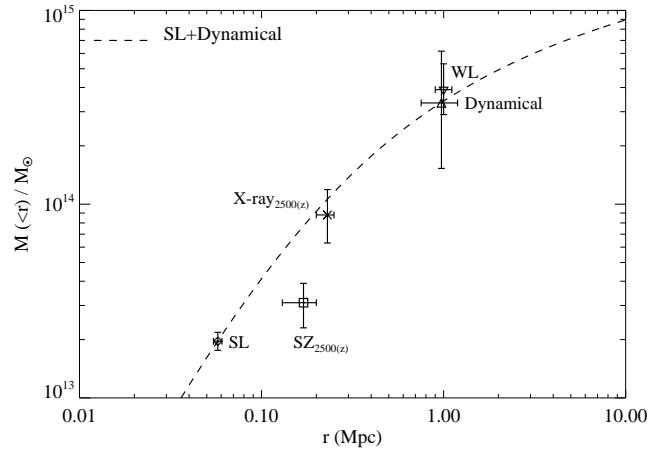}
\end{center}
\caption[Central Region of WARPS1415+36]{\label{fig:nfwprofile}
Comparison of different mass estimates for WARPS J1415+36.  The strong lensing, weak 
lensing and dynamical mass estimates 
presented in this work are labeled above.
$M_{2500(z)}$ from the X-ray \citep{maughan06a} and SZ data \citep{muchovej07a}
are also shown.  The dashed line is an NFW profile fit 
to the strong lensing mass at $r_{E}$=0.0575 Mpc and the dynamical mass at $r_{200}$=0.97 Mpc
($c = 4.96^{+1.81}_{-1.03}$ and $r_{s} = 0.179^{+0.078}_{-0.055}$ Mpc).  
}
\end{figure}

The dynamical mass is estimated from the redshifts of the 21 cluster members 
with $Q>1$ in Table~\ref{tab:members} using the virial scaling relation from 
the simulations of \citet{evrard08a},
\begin{equation}
\sigma_{\rm DM}(M, z)=\sigma_{\rm DM, 15} \left ({h(z)M_{200}\over{10^{15}M_{\odot}}} \right) ^{\alpha}
\label{eqno1}
\end{equation}
where $\sigma_{\rm DM}(M, z)$ is the dark matter velocity dispersion, 
\textit{h}(\textit{z}) = $H (z)/100\ \rm km\ s^{-1}$, and $M_{200}$ is the mass
within $r_{200}$ (the radius of a sphere where the mean density is 200 times the critical density 
at the redshift of the cluster). 
The constants $\sigma_{\rm DM, 15} = 1082.9 \pm 4.0\ \rm km\ s^{-1}$ and $\alpha = 0.3361 \pm 0.0026$
are determined from the simulations of \citet{evrard08a}.
Assuming that the velocity bias 
$b_{v} = \sigma_{\rm v, gal}/\sigma_{\rm DM}=1$,
we find $M_{200} = 3.33^{+2.83}_{-1.80} \times 10^{14} M_{\odot}$ and 
$r_{200} = 0.97 \pm 0.22$ Mpc.  Thus the mass to light ratio is $\sim 114\ M_{\odot}/L_{B\odot}$.  
In Figure~\ref{fig:nfwprofile}, we fit an NFW profile 
\citep*{navarro97a}
to the total cluster mass using the strong lensing and dynamical mass estimates. 
The model has a concentration parameter $c = 4.96^{+1.81}_{-1.03}$ 
and a core radius $r_{s} = 0.179^{+0.078}_{-0.055}$ Mpc, where uncertainties are computed
via Monte Carlo simulation.   

For a simple test of the mass-observable relation in WARPS J1415.1+36,
we compare our strong lensing and dynamical estimates and a model of the mass profile to
estimates of weak lensing mass, X-ray mass, and SZ mass in Figure~\ref{fig:nfwprofile}.
Using the same deep ACS images, Jee et al.\ (2009, in preparation) derive 
the weak lensing mass assuming an NFW profile with a mass-concentration relation from
\citet{bullock01a}.
The X-ray mass $M_{2500(z)}$ \citep{maughan06a} is determined from a best fit two-dimensional 
elliptical $\beta$-model analyzed at $r_{2500(z)}$.
The SZ value of $M_{2500(z)}$ is derived from a spherical, isothermal $\beta$-model
fit to 30 GHz data from the SZ Array \citep{muchovej07a}.
In the X-ray and the SZ mass estimate,
$\Delta(z)$ is a redshift dependent density contrast as described in e.g. \citet{maughan06a}.
Both the weak lensing mass and the X-ray mass are consistent with our results.
However the SZ mass is lower than the prediction of our model by roughly a factor of two. 
In \citet{muchovej07a}, it is noted that the SZ mass estimate is marginally lower than 
the X-ray value, likely due to 
low S/N in the SZ measurements, uncertainties in the absolute calibration, and possible systematic 
errors from the model assumptions.
Additional observations with the SZA at 30 GHz and 90 GHz should improve the constraints
on the SZ mass estimate and shed light on the apparent discrepancy.

Using numerical simulations, \citet{dolag04a} predict the dependence of the 
average halo concentration parameter on the cluster redshift and virial mass
\begin{equation}
c(M, z) = {c_{0}\over{1+z}} \left ({M\over{M_{*}}}\right)^{\gamma}.
\label{eqn:cparam}
\end{equation}
\citet{shaw06a} find the best fit parameters to be $c_{0}=6.47\pm0.03$ and $\gamma=-0.12\pm0.03$
from $N$-body simulations.  Here $M$ is the virial mass and $M_{*} = 11.0\times 10^{14} M_{\odot}$
\citep{shaw06a}.  
If we assume the dynamical mass ($M_{200}$=$3.33\times 10^{14} M_{\odot}$) 
to be the virial mass, Equation~\ref{eqn:cparam} predicts the average
concentration to be $c=3.70^{+0.62}_{-0.34}$$^{+1.9}_{-1.3}$, 
where the first set of errors are from the errors
in the best fit parameters ($c_{0}$ and $\gamma$) and the second set is from the scatter about 
the average \citep{bullock01a}.  Our concentration of $c$ = 4.96 
is higher than the predicted value of $c$ but is well within its errors.

We now examine the strong lensing arcs B, C, and D around cluster member \#13.
The radius of the Einstein ring for the lensing galaxy is estimated by the distance 
between the galaxy and arc C, 
$0.65{\2pr}^{+0.35\2pr}_{-0.30\2pr}$.  The upper and lower bounds are set by the distances from the 
lensing galaxy to arcs D and B, respectively.  Assuming that
the deflection angle due to the cluster potential does not vary appreciably
for arcs B, C and D, the mass enclosed within the Einstein radius is estimated to be 
\(1.65^{+2.25}_{-1.17}\times10^{11}\) \(M_{\odot}\).

\section{Summary}\label{sec:summary}

The galaxy cluster WARPS J1415+36 at \(\textit{z}_{lens}=1.026\) is one of the most distant strong 
lensing clusters reported \citep{thompson01a, gladders03a, gilbank08a}.  Although the SZ mass 
estimate is not consistent with the rest of the data for this cluster,
this work suggests that the mass-observable relations derived from numerical simulations
and observations of strong lensing, weak lensing, dynamical mass, and X-ray mass
still hold for clusters at redshifts at $z \sim 1$.

Recently \citet{gilbank08a} report multi-probe mass estimates for the strong lensing 
cluster, RCS2319, at $z=0.91$.  Future surveys will soon reveal many new high redshift, strong 
lensing systems and provide deep, multiwavelength images of these clusters.
With the corresponding improvements in mass-observable relations in the decelerating regime,
galaxy clusters will soon become an even more effective tool for cosmology.


Financial support for this work was provided by NASA through program GO-10496
from the Space Telescope Science Institute, which is operated by AURA,
Inc., under NASA contract NAS 5-26555.  This work was also supported
in part by the Director, Office of Science, Office of High Energy and
Nuclear Physics, of the U.S. Department of Energy under Contract
No. AC02-05CH11231, as well as a JSPS core-to-core program
``International Research Network for Dark Energy" and by JSPS research
grant 20040003.
Additional support was provided by a scientific research grant (15204012)
by the Ministry of Education, Culture, Sports, Science, and Technology of Japan.
TM and YI were financially supported by the JSPS Research Fellowship.  
Support for MB was provided by the W. M. Keck Foundation.  The work of LAM 
was carried out at the Jet Propulsion Laboratory, California Institute
of Technology, under a contract with NASA.

Subaru observations were collected at Subaru Telescope, which
is operated by the National Astronomical Observatory of Japan.
The authors wish to recognize and acknowledge the very significant
cultural role and reverence that the summit of Mauna
Kea has always had within the indigenous Hawaiian community.
We are most fortunate to have the opportunity to conduct observations
from this superb mountain.


\end{document}